# Application of Genetic Algorithm on Quality Graded Networks for Intelligent Routing

T. R. Gopalakrishnan Nair
ARAMCO Endowed Chair- Technology, PMU, KSA
Advanced Networking Research Group
VP, Research, Research and Industry Incubation Center
(RIIC), Dayananda Sagar Institutions,
Bangalore - 560078, India
e-mail: trgnair@ieee.org, www.trgnair.org

Kavitha Sooda
Advanced Networking Research Group (RIIC), DSI
and Asst. Professor, Dept. of CSE
Nitte Meenakshi Institute of Technology
Bangalore - 560064, India
e-mail: kavithasooda@gmail.com

*Abstract*—**In the past decade, significant research has been carried out for realizing intelligent network routing using advertisement, position and near-optimum node selection schemes. In this paper, a grade-based two-level node selection method along with genetic algorithm (GA) is proposed for realizing an efficient routing scheme. This method assumes that the nodes are intelligent and that there exists a knowledge base about the environment in their local memory. There are two levels for approaching the effective route selection process through grading. At the first level, grade-based selection is applied and at the second level, the optimum path is explored using GA. The simulation has been carried out on different topological structures, and a significant reduction in time is achieved for determining the optimal path through this method compared to the non-graded networks.**

## I. INTRODUCTION

Application of intelligence in to Networking has been one of the fast-changing areas of research, with a high impact on the society. One of the major demands of a network is the capacity to support the vastly increasing level of applications that need to co-exist with the existing infrastructure as well as the future models. The existing infrastructure lacks the understanding and coordination of numerous applications, which leads to various anomalies. In order to meet the demand for improved network, the nodes need to be intelligent and capable of making decisions on their own. The nodes would also need to be aware of the network environment in order to become accurate in deriving the required result. This will enable the network to possess the ability to reason out, learn and remember.

Some aspects of such intelligence have been dealt with in cognitive network [1, 2] and autonomic networking [3]. Attempts are made to formulate networks based on the heuristic algorithm, bio-inspired computing, evolutionary algorithm and the human immune system. These aspects are highly prevalent in research and are applied for making the network intelligent. In the field of networking, application of the above-mentioned areas has its significance in communication, QoS metric, security aspects, application software, and layer management of protocol architecture.

This paper presents the implementation of a novel idea of applying intelligence, which enables the nodes to make decisions and carry out the routing of packets. This intelligence at the node level makes the routing efficient and better. Results show that the grading approach along with genetic algorithm (GA) has achieved the required routing path with promising speed.

The rest of this paper is organized as follows; in Section II related literature is described and in the next section the proposed method is discussed in detail. Simulation results and analysis are presented in Section IV. Conclusion of the paper is provided in the last section.

## II. RELATED WORK

With the increasing number of users of the Internet, it is a challenging task to provide effective connectivity with proper bandwidth to individual users. Static routing and dynamic routing approaches are the main modes of routing in a network. The routing decisions in the static case are merely based on the table information provided at the node level. It has no awareness of the environment around it. This is where autonomic networks (AN) with applied intelligence can play a role, and many researchers have been working in this area for the past few years. Motorola, IBM and a few other organizations are interested in this domain. Game theory, probability, linear programming, artificial intelligence, genetic algorithm, evolutionary algorithm, artificial immune system and many more stochastic approaches have been applied to derive the knowledge from the network.

The ability to be aware of network operation and adjusting the parameters according to the needs of the scenario allow the network to be cognitive. For cognition to run successfully, network elements such as routers, base station and memory elements are required, which help the host to reconfigure the network as necessary. The cognitive network behaves in reference to an active network, but also includes adaptation and learning technique [4], which makes the process different (Fig. 1).

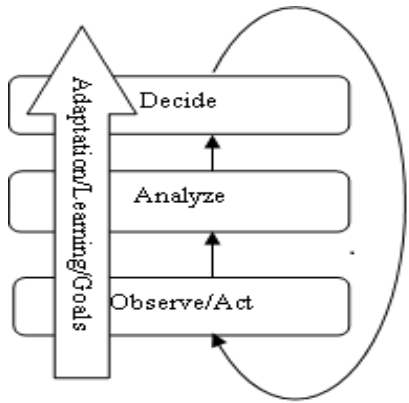

Figure 1. Cognitive cycle which helps in the learning process. The goals are always remembered while learning.

Cognition was conceptualized by Mitola [5], and later the idea of a feedback loop [6] was derived from it. With many advances achieved in the field of cognition, the autonomic approach became more accepted to solve many issues related to a network. This approach was also applied to business aspects, technical issues and different functionality at layers. Mitola [5] states that in order to become autonomic, it is not sufficient if the system possessed the self-* properties. Along with self-* learning, the system needs to be environment-aware. This awareness adds the capability to gain knowledge, monitor and adjust to the current scenario. It is a challenging task to have a representation of a system with the knowledge to form the autonomic system. Understanding the system capability, the functional requirements of the user and mapping of the requirement to the system design along with the limitation of the system itself require a lot of in-depth surveys of the system as a whole [6].

*Genetic Algorithms*

Genetic algorithm is a category of evolutionary algorithm [7]. The algorithm proposed in this paper is based on the concept used in it. It consists of a population with chromosomes that represent the possible solutions for the given problem space. With the developments that take place in the population based on the fitness function evaluation, chromosomes are given a chance to reproduce and be closer to the solution. The reproduction techniques involved are crossover and mutation. This process is repeated for a desired number of times to obtain the optimal solution. The importance of GA lies in the parallel working of different populations. This helps one to explore the complete problem space in all directions. GA is well suited in cases where the problem space is huge and time taken to search is exhaustive as discussed in [8]. It does not require any previous knowledge to obtain the solution.

The following scenario illustrates the operation of crossover and mutation:

Crossover example:
   a. 1-point example:
        Parent1: **1, 3, 5, 7, 2, 4, 6, 8**
        Parent2: **1, 1, 2, 2, 4, 5, 7, 6**
        Random choice: $k = 5$
        Child: **1, 3, 5, 7, 2, 5, 7, 6**
   b. 2-point example:
        Parent1: **1, 3, 5, 7, 2, 4, 6, 8**
        Parent2: **1, 1, 2, 2, 4, 5, 7, 6**
        Random choices: $j = 3, k = 5$
        Child: **1, 3, 5, 2, 4, 4, 6, 8**

Mutation example:
   a. Swap mutation example:
        Parent: **1, 3, 5, 7, 2, 4, 6, 8**
        Random choices: $i = 3; j = 6$
        Child: **1, 3, *4*, 7, 2, *5*, 6, 8**
   b. Adjacent-swap mutation example
        Parent: **1, 3, 5, 7, 2, 4, 6, 8**
        Random choice: $j = 6$
        Child: **1, 3, 5, 7, 2, *6*, *4*, 8**

The fitness function [9] used for the selection of the chromosome is as follows:

$$fj(t) = \frac{Bj(t)}{\sum_{i=0}^{l} Bi(t)} \qquad (1)$$

### III. GRADE SELECTION APPROACH

Grade value estimation method for implementing intelligent routing [9] in the autonomic network is the core focus of research thrust. Grade is like an index; it is made available everywhere and routing will much depend on it. It signifies the quality of the router, which is in fact the knowledge of the environment. The router must be an intelligent entity because it performs different operations based on the environment information. It depends on input, output, load and resource availability.

*Necessity of grading*

Grade largely provides information about the path. All kinds of knowledge about the path make the routing efficient. Then the efficiency depends on the kind of knowledge base that is possessed by the node. Many researchers have defined grade as hop distance [10] and have obtained better results by considering the value of the grade.

As there is a lot of external influence on the node in the autonomic network, the network topology is dynamic in nature. Hence the condition of all nodes must be made available everywhere. This defines the health condition and utility factor of that particular node. Overall network grade in local vicinity can be done by any heuristic algorithm, which shall search and find the best possible nodes that can participate in routing. Grading depends on the intelligent level chosen for operation. When varying level intelligence is considered, defining rules and regulation protocol is challenging. Once these factors are well-defined, multi-grade network can be easily achieved.

A grade value must be calculated in real-time based on observed factors. The challenge lies in identifying the right indicator for the input of the grade based function. The choice of the indicator depends on various factors like distributed environment, reliability in the presence of external perturbations, internal perturbations and resource availability. One way to solve this issue is to study the topological network and simply work on pre-defined training cases. Research has proved that a run-time stochastic approach has better results in obtaining the training set. If any deviation is detected, there is an appropriate action taken to overcome it. However, learning happens there at system startup, which need not be suitable for the application wherein the population is generated randomly later. The challenge also lies in the protocol, which depends on the nodes located at different parts. Collecting information cumulatively at one location to find the best possible indicator is a difficult task. A lot of research scope is yet to be resolved in this direction. A few topics have been dealt with in [11].

Game theory, probability, linear programming, evolutionary algorithm, genetic algorithm artificial immune system, artificial intelligence and many more stochastic approaches have been applied to achieve the knowledge from the network. This work is based on the grade function, which takes varying parameters depending on the environment condition. Initially, the work can be carried out for a fixed parameter to understand the working principle of the grade function and to test the efficiency. The grade function can be selected based on,

- elasticity of application traffic
- human psychology based on mean opinion score
- resource allocation efficiency
- fairness of different shapes of utility function, which leads to optimal resource allocation
- Rate, reliability, delay, jitter, power level
- Congestion level, energy efficiency, network lifetime, collective estimation error.

*A. Queueing systems*

A queueing model is used for mathematically analyzing the queueing behavior. We can obtain many steady state performance measures, which include: average number in the queue, probability of finding the system in a particular state and a queue being full or empty and statistical distribution of the number of queues. These performance measures are important to be considered with respect to the service offered by the link. Analysis of such a queueing model will help us to identify the issue and the impact of the changes to be assessed. Queueing model is a stochastic model that represents the probability that a queueing system will be found in a particular configuration. It is represented using Kendall's notation. The model used for the current work is M/M/1 [12]. It stands for Markovian inter-arrival time along with exponential service time with a single server. This model represents the steady state of the system under consideration. The exponential distribution time describes the event's occurrence and independence at a constant average rate. The Poisson statistical model is a generally accepted tool to predict end user behaviour. Little's formula for M/M/1 queue for mean number of jobs in the system is given by,

$$E(n) = \frac{\rho}{1-\rho} * \frac{1}{\lambda} \qquad (2)$$

Here $\rho$ represents the traffic intensity and $\lambda$ represents the mean arrival rate of the message flow.

*B. Implementation of grade*

This paper is based on the following list of parameters which helps in grading according to priority,

- network lifetime (NL)
- node density (ND)
- traffic congestion (TC)
- resource allocation (RA)
- delay of the packet arrival
- bandwidth availability

*C. Level-1 and Level-2 operations*

*Level-1* is applied region-wise and the goal is to achieve favorable routing based on selected attributes. The values obtained from *Level-1* must be able to eliminate the non-production node. Non-production nodes are those that come in the pitfall of congestion and possess less resource availability. These nodes must be identified by the algorithm that defines the gradient from most non-productive to productive nodes in a homogeneous network. This is identified by assigning a grade value from -3 to +3, i.e. it signifies the productivity value of the node. At this point, we are able to calibrate the routing process region-wise. The algorithm requires proactive decision making on the output obtained. This is because many paths exist to reach the destination, and we have to choose the most optimal path. Once the gradient value has been calculated, it can be made available as pervasive information packets to all the other nodes to obtain the optimal path constitution.

Now the graded function, i.e., *Level-2*, considers the calculated values as its input of *level-1*. The output of this function defines the route availability for the set of nodes considered. This shall be calculated for all the available paths leading towards the destination node. The mean value of the gradient in the grade function shows the success level of operation of the network.

*Level-1 operation*

Assume top three attributes are selected for every region based on priority assigned (Fig. 2).

**Step 1:** The top three priority nodes are selected.
**Step 2:** Select the nodes which are nearing to the optimal (relaxation of ± 2).

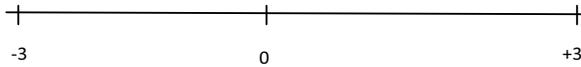

Figure 2. Scale of prioritization assigned to the nodes after observation. Here -3 represents no network lifetime, 0 represents most optimal node and +3 represents nodes which can be considered.

**Step 3:** Selection is done for the nodes that would satisfy the relaxation range from 0 to 2.
**Step 4:** Build a connectivity graph for making sure connectivity exists between regions and apply second level grade to find the optimal path (Fig. 3).

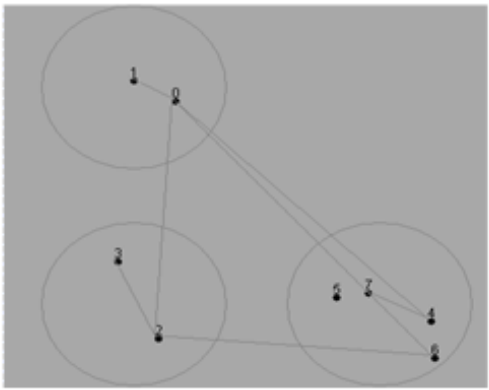

Figure 3. Region based network topology.

Assessing the utility parameter increases the grade value. This approach helps in considering the output for routing.

*Level-2 operation*

The topology obtained from *Level-1* operation is considered as input data here. GA selection mechanism is applied based on the bandwidth availability for the path determination.

**Step 1:** Consider all possible paths from source to destination as input set.
**Step 2:** Execute the following steps for *K* generations:
   a. Select one chromosome based on elitism technique.
   b. Select the remaining of the chromosome based on 90% crossover and 10% mutation.
   c. Generate *N* such chromosomes for each generation.
   d. Calculate the fitness value of each one of the chromosomes. The chromosome whose fitness value is above 0.9 is carried forward to the next generation.

**Step 3:** Select the chromosome whose fitness value is above 0.9.

Here the fitness is calculated based on the bandwidth available at the node.

*Design aspects*

The design involves the generation of an input model, priority model, gradient algorithm and knowledge base. The input model is based on the M/M/1 technique. The priority model is obtained as shown in Fig.4. The gradient algorithm is the *Level-1* operation as discussed earlier. The knowledge base contains the nodes and paths with good condition. Six parameters are considered to obtain the required result. The following parameters are made available in a vector:
   a. Resource allocated
   b. Network lifetime
   c. Bandwidth at nodes

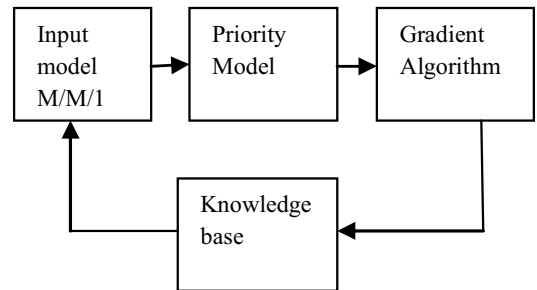

Figure 4. Flow diagram of the implementation

Fig. 4 depicts the input M/M/1 model after which we next apply the priority model from which certain number of nodes gets selected by *Level-1*. Next we apply *Level-2 operation*, which is the gradient algorithm by which we determine the optimal path. This is the knowledge gained by the network and is fed to the knowledge base.

*Delay*, *congestion* and *Node Density* are calculated. *Delay* is derived from service rate, arrival rate and capacity. *Congestion* is derived from the expected data rate at the nodes. *Node Density* to be calculated is based on in-degree of the topology set-up. Information about the connectivity must be made available in a file. For the first level of node selection we apply the first level selection method. The nodes are prioritized based on local observation. Prioritizing is performed based on ontological reasoning [16] as shown in Fig. 5.

From this top three priority nodes are selected. Checking is carried out to see whether delay has reduced and resource allocation is made available. Verification is also performed for traffic congestion. If all three parameters are satisfied then nodes can participate for the next level of selection.

```
IF (NL) {
    IF (ND < 5) {
        IF (TC does not exist) {
            IF (RA) {
                IF (Delay does not exist) {
                    P=1;
                ELSE
                    P=2;
            ELSE
                P=3;
        ELSE
            P=4
    ELSE
        P=5;
ELSE
    P=6;
```
Figure 5. Priority model

Now we categorize the favorable and non-favorable nodes and look for connectivity between the regions. Select the source and destination node. The next step involves checking the bandwidth availability and applying GA with quality grading of nodes. These procedures lead to the realization of optimal path, as discussed earlier.

*Delay calculation*

We assume that the node location, external traffic requirements $\gamma_{ik}$, channel cost $d_i(C_i)$, the constants $D$, $\mu$ and the flow ($\lambda i$) are given and feasible. Thus the Delay $T$ at the node is given by,

$$T = \sum_{i=1}^{M} \frac{\lambda i}{\gamma} \left[ \frac{1}{\mu C i - \lambda i} \right]$$

(3)

The average rate of message flow $\lambda_i$, on the $i^{th}$ channel is equal to the sum of the average message flow rate of all paths that traverse this channel. The traffic entering the network from the external sources forms a Poisson's process with a mean $\gamma_{jk}$ (messages per second) for those messages originating at node *j* and destined for node *k*. Therefore the total external traffic entering and leaving the network are considered to be equal to $\lambda_i$.

## IV. SIMULATION RESULTS

The topology was set up using region-based design approach. A random topology was set-up and tested for different topological structures. Initially, the path selection took place region-wise. Later the regions were considered for connectivity based on the initial setup. Six parameters are considered for the simulation. Five parameters were considered by *Level-1* and the sixth parameter was considered by *Level-2* approach which involves GA. Table 1 describes how the values were obtained.

TABLE I. INPUT PARAMETERS

| Parameter list | Description |
|---|---|
| Bandwidth at nodes | Random number |
| Congestion | Calculated based on bandwidth available and current traffic |
| Delay | Calculated based on M/M/1 model |
| Network lifetime | Random number |
| Node density | Number of in-degree to the node |
| Resource allocated | Random number |

Now *Level-1* operation is applied to obtain a concise network topology which would have the knowledge of the environment where the topology belongs. Most of the learning phase would have been carried out at this stage. Here the nodes which have the maximum lifetime, least congestion and nodes with resource availability would have been selected. The rest would be considered as non-productive nodes and would not participate for the next level grade selection scheme.

Now the nodes which would have been selected by *Level-1* operation would go as input for *Level-2* operation. Here the bandwidth availability at the node level is considered for the selection. This value would have been randomly determined and the simulation results are obtained.

Table 2 gives a comparison of optimal path determination of GA with quality grading of nodes and GA without quality grading of nodes for one of the random runs of the model. Here, the GA with quality grading of nodes determines the path based on the quality of the node, thus making it more reliable for best path determination. In GA without quality grading of nodes, the few nodes that were selected were either congested or with less resource, and a few led to congestion. Thus, determining the optimal path took more time.

Table 3 shows that the GA with quality grading of nodes was better in terms of node density, delay, congestion and route length as the computation was based on state aware network. The results show that better and more deterministic paths can be obtained by the method proposed.

Fig. 6 shows that the number of nodes selected by the GA with quality grading of nodes was lesser and more reliable than the GA without quality grading of nodes.

TABLE II. OUTPUT ANALYSIS

| Total nodes considered | GA with quality grading of nodes | | GA without quality grading of nodes | |
|---|---|---|---|---|
| | Number of nodes selected | Route length | Number of nodes selected | Route length |
| 4 | 4 | 2 | 2 | 1 |
| 8 | 6 | 4 | 5 | 3 |
| 16 | 12 | 6 | 11 | 5 |
| 32 | 24 | 9 | 24 | 10 |
| 64 | 33 | 12 | 28 | 12 |
| 128 | 38 | 6 | 35 | 7 |
| 256 | 42 | 9 | 34 | 10 |

TABLE III. COMPARISON OF THE TWO METHODS

| Algorithm / Issues | GA with quality grading of nodes | GA with quality grading of nodes |
|---|---|---|
| Level-1 parameters | No | Yes |
| Knowledge Base | No | Yes |
| Shortest path | 90% | 90% |
| Intelligence aspects | No | Yes |

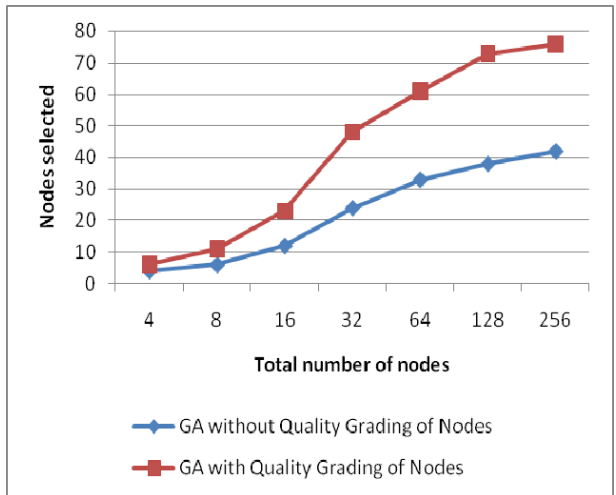

Figure 6. The graph shows the nodes selected by the two methods.

## V. CONCLUSION

The results obtained show a significant improvement in the convergence of optimal path by using GA with quality grading of nodes. The comparative results of the two approaches show that modified GA selected only the best nodes for path determination when compared to GA without the awareness of the state of the network.

Further, the algorithms may be improved with multi-parameters [13] that need to be considered to assess the grade function. This can be a homogeneous grade or a high grade network that will be organizing collective information at the nodes, where decision can be derived by intelligent arbitration.